# Spin-Orbital Coupling in All-Inorganic Metal-Halide Perovskites: the Hidden Force that Matters

Pradeep Raja Anandan[1], Muhammad Nadeem[2], Chun-Ho Lin[1], Simrjit Singh[1,3], Xinwei Guan[1], Jiyun Kim[1], Shamim Shahroki[1], Md Zahidur Rahaman[1], Xun Geng[1], Jing-Kai Huang[1], Hien Nguyen[1], Hanlin Hu[4], Pankaj Sharma[1,5,6], Jan Seidel[1], Xiaolin Wang[2,7,a)], and Tom Wu[1,8,a)]

[1]School of Materials Science and Engineering, Faculty of Science, University of New South Wales, Sydney 2052, Australia

[2]Institute for Superconducting and Electronic Materials, Australian Institute for Innovative Materials, University of Wollongong, Wollongong, NSW 2500, Australia

[3]Department of Applied Physics and Science Education, Eindhoven University of Technology, Eindhoven 5600 MB, The Netherlands.

[4]Hoffmann Institute of Advanced Materials, Shenzhen Polytechnic, 7098 Liuxian Boulevard, Shenzhen 518055, China

[5]Australian Research Council Center of Excellence in Future Low Energy Electronics Technologies, UNSW, Sydney, NSW 2052, Australia

[6]College of Science and Engineering, Flinders University, Bedford Park, Adelaide, SA 5042, Australia

[7]Australian Research Council Center of Excellence in Future Low Energy Electronics Technologies, University of Wollongong, Wollongong, NSW 2500, Australia

[8]Department of Applied Physics, The Hong Kong Polytechnic University, Kowloon, Hong Kong 999077, China

[a)]Authors to whom correspondence should be addressed: xiaolin@uow.edu.au, tom.wu@unsw.edu.au

**Abstract:**

Highlighted with improved long-term thermal and environmental stability, all-inorganic metal halide perovskites exhibit tuneable physical properties, cost-effective synthesis, and satisfactory optoelectronic performance, attracting increasing research interests worldwide. However, a less explored feature of these materials is their strong spin-orbit coupling (SOC), which is the hidden force influencing not only band structure but also properties including magnetoresistance, spin lifetime and singlet-triplet splitting. This review provides an overview of the fundamental aspects and the latest progress of the SOC and debate regarding Rashba effects in all-inorganic metal halide perovskites, providing critical insights into the physical phenomena and potential applications. Meanwhile, crystal structures and photophysics of all-inorganic perovskite are discussed in the context of SOC, along with the related experimental and characterization techniques. Furthermore, a recent understanding of the band topology in the all-inorganic halide perovskites is introduced to push the boundary even further for the novel applications of all-inorganic halide perovskites. Finally, an outlook is given on the potential directions of breakthroughs *via* leveraging the SOC in halide perovskites.

# 1. Introduction:

## 1.1 Basic physics of SOC and Rashba effects

Spin-orbit coupling (SOC) was first reported in semiconductors with zinc-blende structures.[1] Even though these materials are non-magnetic, they exhibit spin splitting in the conduction band (CB) with no centre of inversion. Paramagnetic semiconductors may exhibit spin-related interactions, such as the spin Hall effect, which was theoretically proposed by D'yaknov and Perel in 1971 and later experimentally proved by Kato *et al.*.[2,3] The SOC phenomenon is important for the spin Hall effect due to the impurity scattering, which leads to spin accumulation at the edges.[3] In semiconductors, an electron with momentum p in an electric field (E) experiences a magnetic field ($B_{eff}$) in its rest frame of reference:

$$B_{eff} \sim E \times p/mc^2, \qquad (1)$$

where m is the mass of an electron and c is the speed of light.

In relativistic framework, the SOC manipulates the orbital motion of the electron with the Hamiltonian matrix:

$$\hat{H}_{SO} \sim \mu_B (E \times p) \cdot \sigma/mc^2, \qquad (2)$$

In this equation, where $\mu_B$ is the Bohr magneton, *m* is the electron mass, and *c* is the speed of light. SOC arises due to the interaction between an electron's intrinsic spin angular momentum ($\sigma$) and its orbital angular momentum, which is related to its motion (p) in the presence of an electric field (E). The SOC leads to the momentum-dependent splitting of energy bands and has the unique ability to control the transport properties of semiconductors without any external fields.[4,5]

In the presence of atomic nuclei or other charged particles,[6] the Hamiltonian of SOC can be written as,

$$H_{SO} = \frac{\hbar}{4m_0^2 c^2} \mathbf{p} \cdot \sigma \times (\nabla V_0) \qquad (3)$$

where $V_0$ is the Coulomb potential, and $\sigma = (\sigma_x, \sigma_y, \sigma_z)$ is the vector of Pauli spin matrices. In halide perovskites, SOC becomes significant for electrons near heavy nuclei like Pb, where it depends linearly on $\nabla V_0$ and p.[7]

The Rashba effect results from the splitting of spin energy levels due to broken inversion symmetry and SOC, specifically stemming from structural inversion asymmetry.[8,9] In the case of materials like halide perovskites with even structural symmetry, the breaking of inversion symmetry may occur due to ionic impurities that induce local electric fields. [8] The Rashba energy splitting $\Delta E_{Rashba}$ caused by an electric field ($\hat{E}$) is represented by:[10]

$$\Delta E_{Rashba} = \alpha_R (\hat{E} \times \mathbf{p}) \cdot \hat{z} \cdot \sigma \qquad (4)$$

where $\alpha_R$ is the Rashba parameter and $\hat{z}$ is the unit vector along the z-axis. Clearly, the Rashba effect influences the energy level splitting of charge carriers when an electric field is present.[11,12]

Similar to the Rashba effect, the Dresselhaus effect also results in spin splitting, but with a different spin texture.[7,13] Dresselhaus studied this phenomenon in crystals lacking a centre of inversion symmetry, exhibiting bulk inversion asymmetry (BIA).[14] SOC is responsible for lifting the spin degeneracy of electronic bulk states in such crystals. The Hamiltonian describing Dresselhaus splitting is represented as follows:[15]

$$H_D = D\left[\sigma_x k_x (k_y^2 - k_z^2) + \sigma_y k_y (k_z^2 - k_x^2) + \sigma_z k_z (k_x^2 - k_y^2)\right]. \qquad (5)$$

where $D$ is the Dresselhaus coefficient, $\sigma_x, \sigma_y, \sigma_z$ are Pauli matrices to represent spin states of the electron, and $k_x, k_y, k_z$ are the components of the wavevector ($k$) representing the momentum of charge carriers in the crystal.

**1.2 SOC in semiconductors: from silicon to perovskites**

Semiconductor technology has revolutionized the modern microelectronics and laid the foundation for advanced computing and memory devices.[16,17] Semiconductors with appropriate bandgaps have unique properties of transforming light into electric current and vice versa, enabling a wide range of solid-state optoelectronic applications.[18,19] The bandgap of semiconductors, namely the energy difference between valence band (VB) and CB edges, directly determines their optical properties, and accordingly, semiconductors are classified into direct or indirect ones. Intrinsic silicon is a prototypical indirect semiconductor with a bandgap of 1.1 eV. As shown in **Fig. 1**(a), the minimum of the CB, $X_1$ does not coincide with the maximum of the VB, $\Gamma$.[20] As a result, photo-excited transition across the bandgap must involve interactions with phonons to keep the total energy and momentum conserved.[21]

Additionally, the spin degree of freedom plays a crucial role in semiconductor behaviour. SOC, which couples an electron's spin with its motion, impacts various aspects of semiconductor physics.[22] Silicon, with its relatively low atomic number, exhibits weaker SOC compared to elements with higher atomic numbers.[23] Understanding the interplay between bandgap properties, spin characteristics, and atomic structures is crucial for semiconductors with strong SOC and their applications in electronics and photonics.[24]

**Fig. 1**(b) depicts the crystal structure of three-dimensional metal halide perovskites, a class of emerging semiconductors with strong SOC.[25] In the chemical formula of $ABX_3$, A stands for organic cations like methylammonium (MA) and formamidinium (FA) or inorganic counterparts like cesium (Cs); B is a metal cation, such as lead ($Pb^{2+}$) and tin ($Sn^{2+}$); and X encompasses halide ions, e.g., iodide ($I^-$), bromide ($Br^-$), and chloride ($Cl^-$).[26-31] Metal halide perovskites are promising candidates for a wide array of optoelectronic applications due to their tunable properties and exceptional performance.[32-36]

Halide perovskites have Rashba-type SOC,[8] which leads to unique spin-related effects and profoundly modifies their properties.[37,38] In general, the electronic behaviours of semiconductor materials are largely determined by the extrema points of their CB and VB. The electron dispersion relation, often approximated through the effective-mass approach, follows ($E(k) = \hbar^2 k^2 / 2m^*$), as shown in **Fig. 1**(c)(i).[39] However, the introduction of SOC in non-centrosymmetric compounds brings about an intriguing departure from this simplification. Specifically, in the presence of structural "inversion asymmetry," the previously spin-degenerate parabolic band structure splits into two distinct spin-polarized bands. The dispersion relationship for electrons and holes in these materials takes on the form, $E^{\pm}(k) = (\hbar^2 k^2 / 2m^*) \pm \alpha_R |k|$, where the parameter "$\alpha_R$" signifies the Rashba splitting. This modification engenders the appearance of new extremal points characterized by a momentum offset ($k_0$) and a corresponding energy separation ($E_R$), linked by the relationship $\alpha_R = 2E_R / k_0$, as portrayed in **Fig. 1**(c)(ii). A remarkable outcome of this phenomenon is the emergence of two separate Rashba-split branches, each possessing a distinctive spin orientation.[40]

Experimental evidence suggested the presence of Rashba spin splitting in 3D halide perovskites, exemplified by MAPbI$_3$.[8] However, a debate persists regarding the existence of the Rashba effect in all-inorganic halide perovskites.[41-43] The role of symmetry is pivotal in understanding these effects in perovskite materials. In 2D hybrid perovskites, Maurer et al. established a connection between lattice distortions and SOC, claiming notable Rashba-Dresselhaus splitting.[7] In-plane Pb displacement is claimed to be responsible for the Rashba effect, and specifically, diagonal Pb displacement favours a pure Rashba splitting, while pure Dresselhaus splitting is not observed.[7] In another recent theoretical study the existence of the Rashba-Dresselhaus effect was confirmed in 2D halide perovskites.[44]

Interestingly, the application of pressure can result in indirect to direct bandgap transition in MAPbI$_3$ due to the decrease in the local electric field surrounding the Pb atoms, reducing the Rashba splitting.[45-48] This effect is accompanied by a phase change in MAPbI$_3$ from *I*4/*mcm* to *Imm*2. The carrier lifetime rapidly falls with pressure as the bandgap becomes more direct, and the photoluminescence quantum yield (PLQY) doubles.[49] By pushing the band edge closer to the subgap traps and making these traps even shallower, pressure increases the carrier lifetime, and the Rashba splitting is reduced, resulting in a more direct bandgap of MAPbI$_3$.[50-52]

In 2016, Niesner et al. utilized angle-resolved photoelectron spectroscopy (ARPES) to make a groundbreaking discovery of a massive Rashba splitting in MAPbBr$_3$.[53] Their findings confirmed the stronger SOC in the CB than the VB, facilitating direct and isotropic optical transitions. This initial report sparked considerable interest in the scientific community. In 2020, Sajedi et al. used ARPES to re-examine MAPbBr$_3$, but their result did not support the previous findings of Rashba splitting at the VB maximum.[50] In a report by Becker et al., magneto-optical photoluminescence (MPL) measurements were used to unveil the Rashba effect in the all-inorganic halide perovskite CsPbBr$_3$ nanocrystals (NCs).[54] This Rashba effect is believed to originate from lattice distortions induced by the motion of Cs$^+$ ions or surface-related influences in CsPbBr$_3$ NCs. Importantly, this Rashba effect induces singlet and triplet energy level splitting without the presence of an external magnetic field. However, this report faced challenges when Tamarat et al. employed MPL spectroscopic measurements to reevaluate the findings.[41-43] Their results suggested that the previously observed energy level splitting in weakly confined NCs might predominantly arise from exchange interactions rather than the Rashba effect. In short, more comprehensive experimental efforts on carefully selected samples under well-controlled conditions are warranted to unravel the existence and characteristics of the Rashba effect in halide perovskites.

**1.3 Effect of SOC on the electronic properties of perovskites**

**Fig. 1**(d) shows the band structure of a prototypical cubic all-inorganic perovskite, CsPbBr$_3$, calculated using the Vienna Ab-initio Simulation Package (VASP).[54] The electronic

bandgap is located at the R point in the Brillouin zone. The VB maximum stems largely from the antibonding hybridization of Pb 6s and Br $n$p orbitals, while the CB maximum mainly originates from the Pb 6p orbital, with some contribution from the Br $n$p orbitals.[55] For halide perovskites, the presence of Pb leads to giant SOC, which is believed to contribute to the low carrier recombination rate.[56] In addition, defects, which typically impact charge transport, bring about a distinct facet of properties in halide perovskites due to the Rashba effect.[56-58] It is a general consensus that defects do not significantly influence the recombination rate and carrier lifetime, rendering perovskites highly defect-tolerant.

In hybrid perovskites, B-site ions are known to determine the band structure and influence electronic and optical properties, but A-site ions also introduce a significant impact.[59] Specifically, the MA molecule induces the correlation between spin texture and structural deformations. Interestingly, this phenomenon is not static; the orientation of the MA molecules dynamically changes at room temperature, causing variations in the spin texture. Furthermore, the replacement of MA with FA or Cs decreases the presence of Rashba effect due to their reduced polarity.[60]

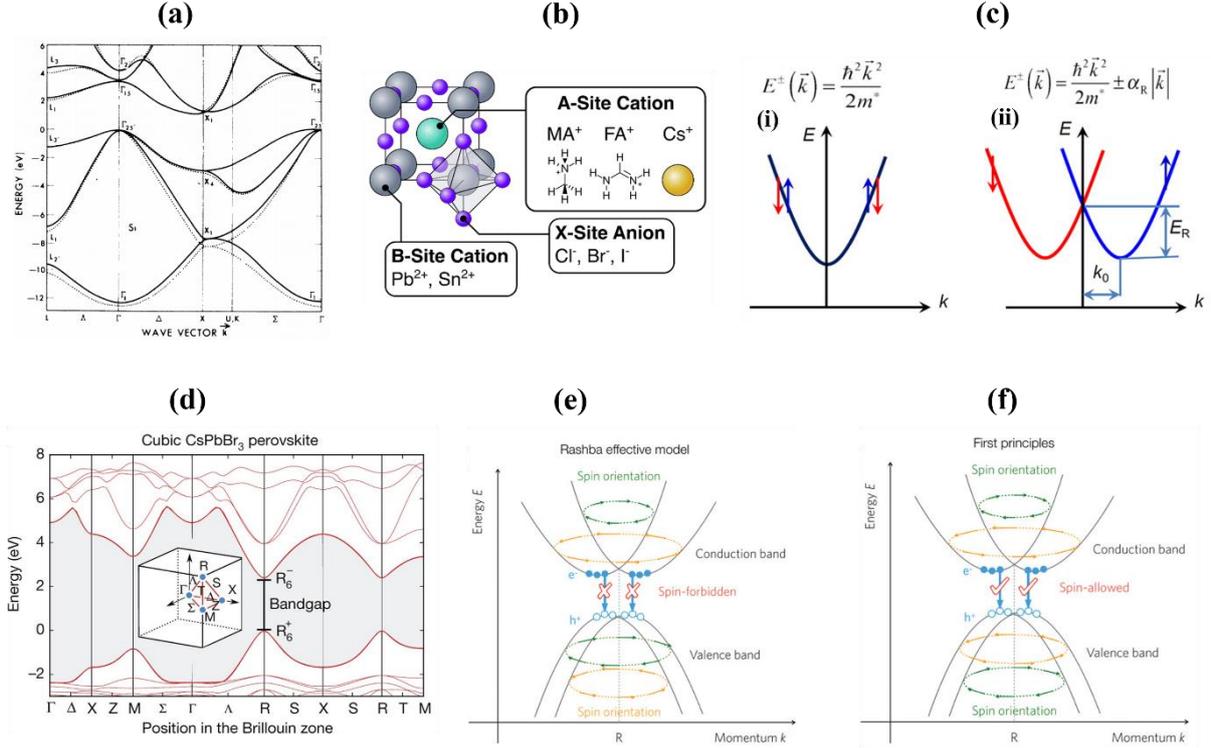

**FIG. 1.** Band structures and SOC of several typical material systems. (a) Partial band structure of silicon. Reproduced with permission. [20] Copyright ©1974 American Physical Society. (b) Schematic representation of a typical halide perovskite crystal structure with formula $ABX_3$. Reproduced with permission. Copyright 2022 Royal Society of Chemistry.[25] (c) (i) Schematic electron dispersion relation of a regular CB that shows a doubly spin-degenerate parabolic band having a single minimum at k=0. (ii) Same as in (i) but subjected to Rashba splitting; two parabolic branches having oppo-site spin sense are formed. The Rashba energy ($E_R$) and momentum offset ($k_0$) are denoted. Reproduced with permission. Copyright 2017 American Association for the Advancement of Science, under a Creative Commons Attribution 4.0 International (CC BY 4.0) license.[39] (d) Calculated band structure of cubic $CsPbBr_3$ perovskite with a bandgap between $R_6^+$ (VB maximum) and $R_6^-$ symmetry (CB minimum). The inset is a cubic crystallite representing the first Brillouin zone. [54] Copyright © 2018 Springer Nature Limited. (e) Schematic of the spin-forbidden transition model.[59] (f) Schematic of the spin-allowed transition obtained from first principles. The green and orange arrows represent the helical spin orientations in different bands. Blue solid and open circles denote photoexcited electrons ($e^-$) and holes ($h^+$), respectively. Reproduced with permission. Copyright © 2018 American Chemical Society.[59]

Charge recombination in halide perovskites is profoundly influenced by the SOC. In the early scenario depicted in **Fig. 1**(e),[59] SOC-induced band splitting near the R point, with opposite spin orientations, causes a mismatch between the lowest CB and the highest VB, reducing direct radiative recombination and extending carrier lifetimes. However, Zhang et al. proposed that direct optical transitions still occur because of the structural distortions and spin texture dynamics induced by the MA molecules, as depicted in **Fig. 1**(f).[59] In other words, the indirect bandgap feature of halide perovskites has a limited influence on the radiative charge recombination. In a recent study, Lu et al. used time-dependent DFT (TD-DFT) and non-adiabatic molecular dynamics (NA-MD) methods to investigate the non-radiative recombination in MAPbI$_3$, both with and without SOC interactions.[61] Their results demonstrated that SOC substantially slows non-radiative e-h recombination, extending the charge carrier lifetime.

It is generally believed that all-inorganic perovskite possesses a SOC effect stronger than most other semiconductors.[62,63] There are some reviews on the importance and advantages of SOC in organic-inorganic hybrid perovskites.[63,64] However, the impact of SOC on all-inorganic halide perovskites, such as CsPbBr$_3$, CsSnBr$_3$ and FAPbBr$_3$, has not been comprehensively reviewed.

This review presents a concise introduction to SOC and its profound impact on the physical properties and optoelectronic applications of all-inorganic halide perovskites. We begin by introducing the fundamental semiconducting and optoelectronic properties of these emerging materials with strong SOC. Subsequently, we discuss the effects of SOC and the Rashba effect on spin lifetime. Moving forward, we explain the experimental techniques used to study spin relaxation and dynamics in all-inorganic halide perovskites. We also discuss how the speculated Rashba effect and quantum confinement influence the light emission in perovskite NCs through modulating the bright and dark states. Finally, we address the recent advancements in understanding the band topology of all-inorganic halide perovskites.

## 2. Basic Properties and SOC of All-Inorganic Halide Perovskites

## 2.1 Basic properties of all-inorganic halide perovskites

The all-inorganic halide perovskite materials have drawn more scientific attention in photovoltaics and optoelectronics because they are more stable under ambient conditions than their organic-inorganic hybrid counterparts.[65,66] $CsPbX_3$ (X = I, Br, Cl) perovskites are one of the promising candidates among all-inorganic halide perovskites due to their stability against moisture, heat, and light. Lasers, solar cells, LEDs, and thermoelectric devices are only a few examples of the many applications based on $CsPbX_3$.[28,54,67] Several reviews have concentrated on the synthesis techniques, surface engineering, phase stability, interface engineering, and composition engineering related to all-inorganic perovskites.[68-70] This work is not concerned with the general properties of inorganic perovskite materials but instead focuses on the debated Rashba effect and the impact of SOC on their optoelectronic properties.

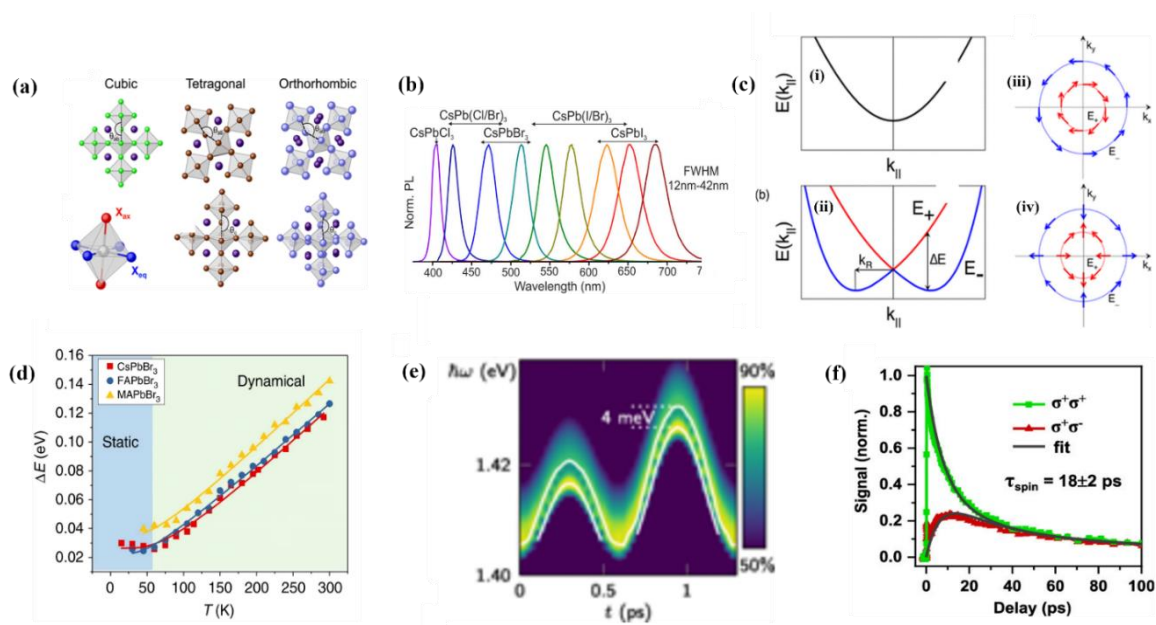

**FIG. 2.** Basic properties and Rashba effect of inorganic metal halide perovskites. (a) Illustration of the crystal structures of $CsPbX_3$ (X = Cl, Br, I), highlighting their varying phases with bond angle. Reproduced with permission. Copyright © 2017 American Chemical Society.[71] (b) PL spectra of $CsPbX_3$ (X=Cl, Br, I). Reproduced with permission. Copyright ©

2015 American Chemical Society.[72] (c) Band dispersion (i) without SOC (ii) with SOC shows degenerated bands with inner ($E_+$) and outer branches ($E_-$). The spin texture represents (iii) pure Rashba and (iv) pure Dresselhaus coupling. Reproduced with permission. Copyright © 2017 American Chemical Society.[73] (d) Temperature-dependent energy differences for $CsPbBr_3$, $FAPbBr_3$, and $MAPbBr_3$. It reveals a static Rashba effect at low temperatures, transitioning to a dynamic Rashba effect above approximately 60 K, driven by thermal-induced octahedral cage deformation. Reproduced with permission. Copyright © 2019 Springer Nature Limited.[74] (e) Peak-dip-hump structure in the PL signal resulted from the dynamic Rashba-Dresselhaus (RD) effect. Reproduced with permission. Copyright ©2021 American Physical Society.[75] (f) CPTA data, from which the spin relaxation time was estimated. Reproduced with permission.[76] Copyright © 2021 American Chemical Society.

The remarkable optoelectronic characteristics exhibited by $CsPbX_3$ (X = Cl, Br, I) perovskites are intricately interwoven with their unique crystalline and electronic structures.[77-80] Their crystal structure is underpinned by a fundamental parameter known as the Goldschmidt tolerance factor.[81] Particularly, $CsPbX_3$ has three distinct crystal structures, i.e., cubic, tetragonal and orthorhombic (**Fig. 2**(a)). The perovskite framework allows for structural tilting of the octahedra. The tilting behaviour involving halide ions situated exclusively in the equatorial plane ($X_{eq}$) leads to a tetragonal structural modification (P4/*mbm*). Alternatively, the tilting phenomenon can encompass both the equatorial plane and the normal axial direction ($X_{ax}$), instigating an orthorhombic structural modification (P*bnm* or equivalently P*nma*). These structural deviations, linked to the Goldschmidt tolerance factor, underlie the unique optoelectronic properties of $CsPbX_3$ perovskites.

Several methods have been developed to synthesize $CsPbX_3$ perovskites in thin-film, nanowire, nanosheet, and NC forms.[82-86] Inorganic halide perovskite thin films have been extensively exploited for photovoltaic applications.[87] The low-dimensional nanostructures, such as nanosheets, nanowires, and NCs, are explored because of their quantum confinement effect, which significantly boosts the exciton binding energy and enhances the PLQY for lighting applications.[88,89] Due to their versatility in modifying their composition, shape, and optical properties, NCs shine out among nanostructures. NCs are typically found to have high surface-to-volume ratios and short radiative lifetimes, making them excellent candidates for catalytic and optical applications.[65,90] Most notably, as demonstrated in **Fig. 2**(b), by adjusting the composition of Cl, Br, and I ions, the optical bandgap of $CsPbX_3$ NCs may be easily controlled, covering the entire visible spectrum from 410 nm to 700 nm. Even under harsh circumstances such as intense X-ray excitation, the colour-tunable emission of $CsPbX_3$ NCs can still be retained, indicating the robustness of these materials.[91]

## 2.2 Importance of SOC in all-inorganic halide perovskites

In **Fig. 2**(c), the band dispersion of a generic material with strong SOC is depicted, where the presence of SOC modifies the band degeneracy, forming inner ($E_+$) and outer ($E_-$) branches.[73] Furthermore, Rashba and Dresselhaus effects lead to different spin textures. The intriguing properties of all-inorganic halide perovskites related to the SOC effect have recently garnered substantial attention. Remarkably, despite being non-magnetic materials, these perovskites exhibit the potential for generating spin-polarized current, thereby promising for future explorations in spin-orbitronics and spintronics.[92]

With the breaking of inversion symmetry, the Rashba effect in perovskites can be either static or dynamic. The static Rashba effect occurs when the Pb framework deviates from its ideal position, leading to the creation of a non-centrosymmetric phase. On the other hand, the dynamic Rashba effect is related to the interaction between the A-site cation and the Pb framework. **Fig. 2**(d) illustrates the static and dynamic Rashba effects at various temperatures ranging from 20 K to 300 K.[93] It is acknowledged that $MAPbBr_3$ experiences a notable static Rashba effect due to the centrosymmetry-breaking MA cations.[94] In $CsPbBr_3$, surface or interface distortions are hypothesized to result in a static Rashba splitting of approximately 26 meV at low temperatures of less than 50 K.[93] Above 60 K, the polar cage of the octahedron undergoes distortion, giving rise to dynamic Rashba splitting.

However, Schlipf et al. challenged this interpretation by demonstrating that dynamic Rashba-Dresselhaus splitting in MAPbI$_3$ halide perovskites can only manifest in the presence of an out-of-equilibrium phonon population, such as that generated by coherent terahertz radiation illumination.[75] The Rashba-Dresselhaus effect is symmetry-forbidden, but the electron-phonon interactions can induce a phonon-assisted, dynamic Rashba-Dresselhaus spin splitting. As shown in Fig. 2(e), the dynamic Rashba-Dresselhaus effect generates a pronounced peak-dip-hump structure in the photoluminescence (PL) signal of MAPbI$_3$.[75] In the case of CsPbBr$_3$, the origin of Rashba splitting is still subject to debate due to the centrosymmetry and the relative position of the triplet state.[41,74]

Ryu et al. extensively investigated the Rashba effects in CsPbBr$_3$ and MAPbBr$_3$ single crystals, employing various techniques.[94] In particular, they unclosed the temperature-dependent behavior of the dynamic Rashba effect, which becomes pronounced at elevated temperatures.[94] Intriguingly, they identified the static Rashba effect in MAPbBr$_3$, manifesting at temperatures below 90 K. This static Rashba effect was attributed to surface reconstruction, induced by the ordered arrangement of MA cations, which introduced structural asymmetry. Conversely, CsPbBr$_3$, characterized by Cs cations and a more symmetric configuration, displayed an absence of Rashba splitting.[94] In a complementary study, Schlipf et al. explored the dynamic Rashba effect in MAPbI$_3$, elucidating its intricate relationship with temperature and lattice vibrations.[75] Their investigation was conducted below 160 K, a regime where MAPbI$_3$ crystallizes into an orthorhombic structure with a centrosymmetric space group (P*nma*). Notably, this crystalline system exhibits 20 infrared-active optical phonons, intimately linked to the deformation of the PbI$_6$ octahedra. The large Rashba energy associated with the CB's spin splitting indicates a significant phonon assisted Rashba effect. Schlipf et al.'s work emphasized the role of phonons in influencing the Rashba effect.[75]

The long spin lifetime of all-inorganic halide perovskites has recently attracted lots of attention. As shown in **Fig. 2**(f), $CsSnBr_3$ NCs demonstrated a spin lifetime of 18 ps at ambient temperature,[76] and its SOC strength is roughly one-third that of its Pb equivalent. **Table I** summarises the spin lifetime of halide perovskites measured via various techniques such as circularly polarized transient absorption (CPTA), Hanle effect, and differential transmission spectroscopy (DTS).[76,95-99]

**Table I.** Summary of spin lifetimes measured in halide perovskites.

| Type of Halide Perovskite | Net Spin lifetime | Temperature | Method | Ref |
|---|---|---|---|---|
| $MAPbI_3$ film | 2.2 ps | RT | CPTA | [96] |
| $MAPbBr_3$ film | 4.3 ps | RT | CPTA | [96] |
| $MAPbBr_3$ film | 240 ps | 77 K | Hanle | [97] |
| $CsPbI_3$ film | 1.3 ps | RT | CPTA | [96] |
| $CsPbBr_3$ film | 3.7 ps | RT | CPTA | [96] |
| $CsPbBr_3$ single crystal | 20 ps | RT | CPTA | [95] |
| $CsPbI_3$ NCs | 3.2 ps | RT | CPTA | [98] |
| $CsPbBr_3$ NCs | 1.9 ps | RT | CPTA | [98] |
| $CsPbI_3$ NCs | 32 ps | 50 K | DTS | [99] |
| $CsSnBr_3$ NCs | 18 ps | RT | CPTA | [76] |
| $FAPbBr_3$ film | 788 ps | 10 K | Hanle | [100] |

Since spin relaxation is a sensitive process, it can vary due to the sample's preparation condition and quality. According to published literature, all-inorganic perovskite $CsPbI_3$ showed spin relaxation of 32 ps at 50 K and 3 ps at 250 K.[99] Recently, Zhao *et al.* reported the exciton spin relaxation of $CsPbBr_3$ is about 20 ps at room temperature, which was the longest among metal halide perovskite materials.[95] The properties such as long exciton spin lifetime and high spin injection rate are considered essential factors in future optoelectronics and spintronic device applications.

For all-inorganic halide perovskite materials, spin-related studies, such as spin relaxation and spin manipulation, are still at the primary stage. In inorganic semiconductors, the spin relaxation mechanisms can be classified into three types: Elliot−Yafet (E-Y),[101] D'yakonov-Perel (D-P),[2] and Bir-Aronov-Pikus (BAP).[102] E-Y and D-P spin relaxation mechanisms are relevant for metal halide perovskite materials. The E-Y mechanism flips the charge carrier's spin by scattering, while in the D-P mechanism, the Rashba effect from structural inversion asymmetry induces spin precession.[2,96,101,102] The understanding of these mechanisms and their interplay with phonon scattering is crucial for manipulating spin states.[96]

Recently, Zhou *et al.* measured the spin lifetime as a function of temperature in $APbI_3$ and $APbBr_3$ (A = $CH_3NH_3$, Cs) halide perovskites.[76] The spin lifetime of $CsPbBr_3$ is reduced from 3.8 ps to 3.2 ps when the temperature decreases from 295 K to 77 K, indicating that the D-P mechanism plays a more significant role in the Br-based halide perovskite. On the other hand, in the case of $CsPbI_3$, the spin lifetime increases from 1.3 ps to 6 ps as the temperature decreases from 300 K to 77 K, indicating that for I-based halide perovskite, the contribution of the E-Y mechanism is dominant.[96] These results provide a promising opportunity to tune the spin-based optoelectronic properties in halide perovskites.

## 2.3. Bright and dark state splitting in all-inorganic halide perovskites

In all-inorganic halide perovskite NCs, the exceptionally efficient light emission is attributed to a distinctive excitonic fine structure mainly composed of triplet states.[103] The presence of a highly emissive triplet state with a significant oscillator strength can be attributed to the complex interaction between electron-hole exchange and SOC of the J = ½ conduction band levels. **Fig. 3**(a) illustrates the band diagram of the exciton state of cubic $CsPbBr_3$ NCs, and the CB and VB feature $\Gamma_4^-$ and $\Gamma_1^+$ symmetry, respectively. Due to SOC, the CB is split into $\Gamma_{8-}$ and $\Gamma_{6-}$ states, and the VB transforms to $\Gamma_6^+$. The exchange interaction between electron and hole results in the formation of a higher bright triplet state ($J=1$) and a lower dark singlet state ($J=0$). [54,104] It was reported that in large NCs of CsPbBr$_3$, the Rashba effect changes the order of bright and dark exciton levels, while the bright-dark level inversion effect is suppressed in small NCs. [103,105,106]

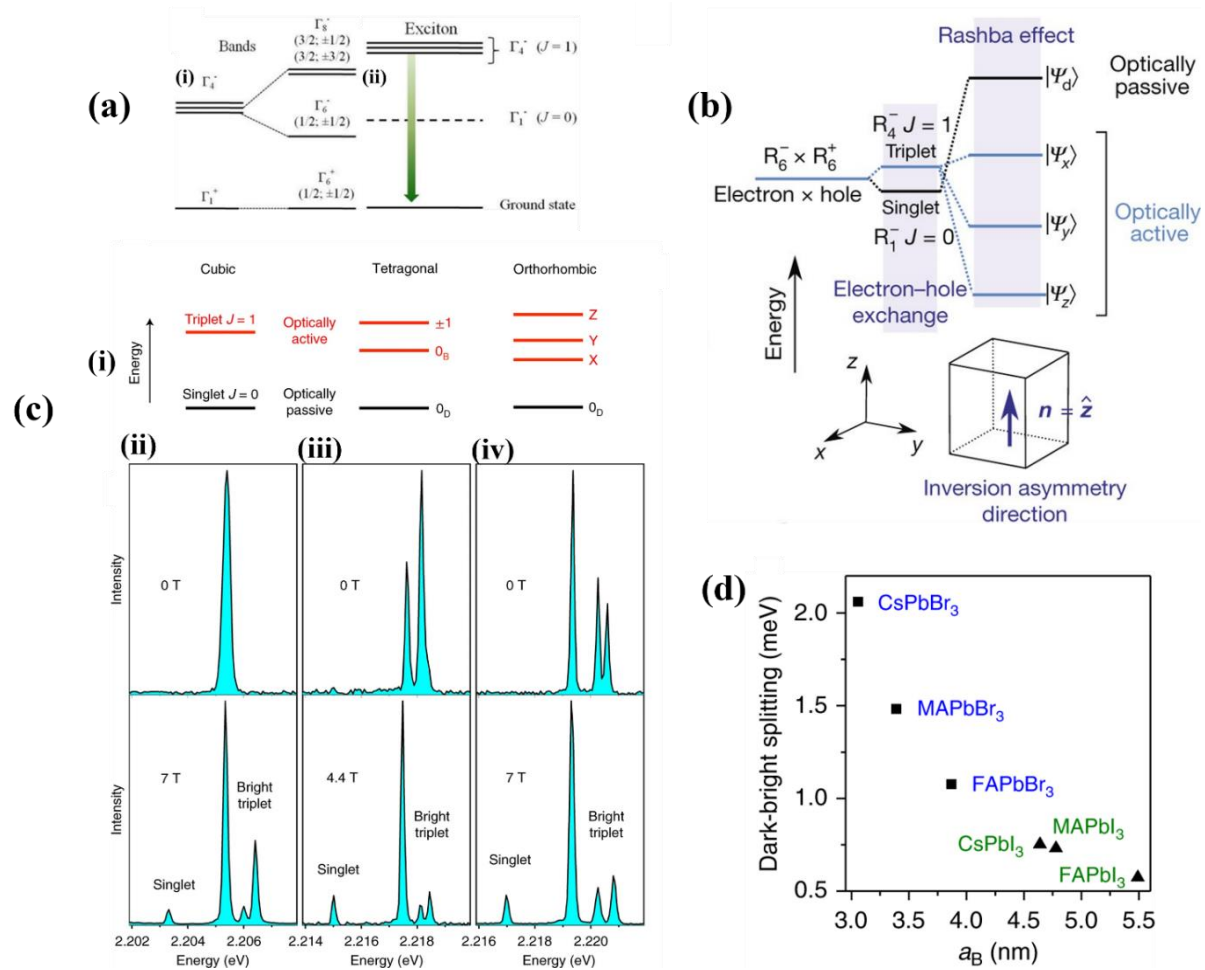

**FIG. 3**. (a) Band diagram illustrating exciton states of cubic CsPbBr$_3$ NCs. (i) CB and VB with $\Gamma_4^-$ and $\Gamma_1^+$ symmetry, respectively. Due to SOC, the CB splits into $\Gamma_{8-}$ and $\Gamma_{6-}$ states. (ii) shows the lowest dark ($\Gamma_{1-}$, $J = 0$) state and the bright ($\Gamma_{4-}$, $J = 1$) exciton state. Reproduced with permission.[104] Copyright © 2017 American Chemical Society. (b) Schematic of the exciton band edge highlighting the electron-hole exchange and the Rashba effect. Reproduced with permission. Copyright © 2018 Springer Nature Limited.[54] (c) (i) Illustration of the band-edge exciton fine structures for three distinct crystal anisotropies, among which the lowest state

corresponds to a dark state, while the others are optically active. (ii–iv) Spectra of three different FAPbBr$_3$ NCs at 4 K under zero magnetic field (upper panels). When subjected to magnetic fields, they transform into four-line spectra (lower panels). Reproduced with permission. Copyright © 2019 Springer Nature Limited.[43] (d) Exchange interactions vs. exciton Bohr radius in a variety of halide perovskites. Reproduced with permission. Copyright © 2020 Springer Nature Limited.[42]

It should be noted that there is still an ongoing debate concerning the relative order of bright and dark excitons in CsPbX$_3$.[41-43,107,108] The short-range exchange in CsPbX$_3$ NCs positions the singlet state below the triplet, but a significant Rashba effect can potentially alter the fine structure, placing the dark singlet exciton above the bright triplet exciton, as shown in **Fig. 3**(b).[54] Only at cryogenic temperatures is the presence of a bright triplet state clearly observed. In contrast, at room temperature there is a mixing of PL excitonic signals due to the small energy difference between the dark singlet state and the bright triplet.[54] Moreover, the emission energy can be impacted by various factors, including composition, phase transition, anisotropy, and lattice disorder.[109,110]

However, Tamarat et al. contradicted the claims in **Fig. 3**(b) and provided an alternative view on the energy level arrangement.[43] The PL spectra from individual FAPbBr$_3$ NCs, particularly sharp zero-phonon lines (ZPLs), revealed bright triplet sublevels, as illustrated in **Fig. 3**(c)(i). The structural variations in **Fig. 3**(c)(ii-iv), featuring one, two, and three lines at 0 T, corresponded to different crystalline phases (cubic, tetragonal, or orthorhombic). An additional emission line emerged from longitudinal optical (LO) phonon sidebands, positioned just below the bright triplet state. This indicated a magnetic brightening phenomenon linked to a low-lying dark exciton state. Importantly, the spectral footprint of the dark singlet exciton also manifested in CsPbI$_3$ under a 7 T external magnetic field.[42] These findings underscore the significant impact of exchange interaction on the fine structure splitting, influenced by quantum confinement, NC shape anisotropy, and electron-hole interaction screening effects.

**Fig. 3**(d) illustrates a substantially stronger exchange interaction in $CsPbX_3$ than the FA and MA counterparts,[42] and the bright triplet splitting suggests a strong crystal field splitting in the orthorhombic crystal structure. Experiments on $CsPbBr_3$ NCs under magnetic fields smaller than 7 T did not reveal the expected 'dark state' due to the narrow energy gap between the dark and bright states. However, exposure to magnetic fields exceeding 10 T induced significant changes in PL, indicating the presence of long-lived energy sublevels.[42,104,111] It is now generally believed that the lowest-energy exciton dark states are located just a few meV below the bright triplet state, resolving the debate on the dark singlet state's position in inorganic lead halide perovskites.[112]

## 3. Spectroscopic Techniques for Investigating the Spin-Related Properties of Halide Perovskites:

The optical techniques to access the crucial properties of halide perovskite materials such as carrier dynamics and spin lifetime often involve PL, reflectivity, and time-resolved photoluminescence (TRPL).[113-115] Techniques such as DTS are primarily used in probing carrier spin relaxation lifetime. Magneto-photocurrent, magneto-electroluminescence (MEL), and MPL are employed to study magnetic field-induced spin mixing of photogenerated e-h pairs.[58,104,116-118] The ultrafast spectroscopy technique has been widely applied for studying quantum well structure in the past,[119] which possesses an excellent time resolution of 100 fs. Spin mixing will modulate the spin population, resulting in different recombination rates and changes in PL/electroluminescence emission intensity.[117] Other techniques can also be combined with TRPL to study the spin-related dynamics at low and room temperatures. This section will extensively cover these techniques applied in probing the spin dynamics of halide perovskites.

## 3.1 Optical techniques

Optical techniques are commonly utilized because they are fast and non-invasive. PL is a primary technique widely applied in perovskite sample characterization, where a chopped, pulsed or continuous-wave (CW) laser beam,[120] is used to excite charge carriers, providing an ideal platform for exploring fundamental material properties, such as ion migration, sample degradation, and phase transition in halide perovskite.[121-126]

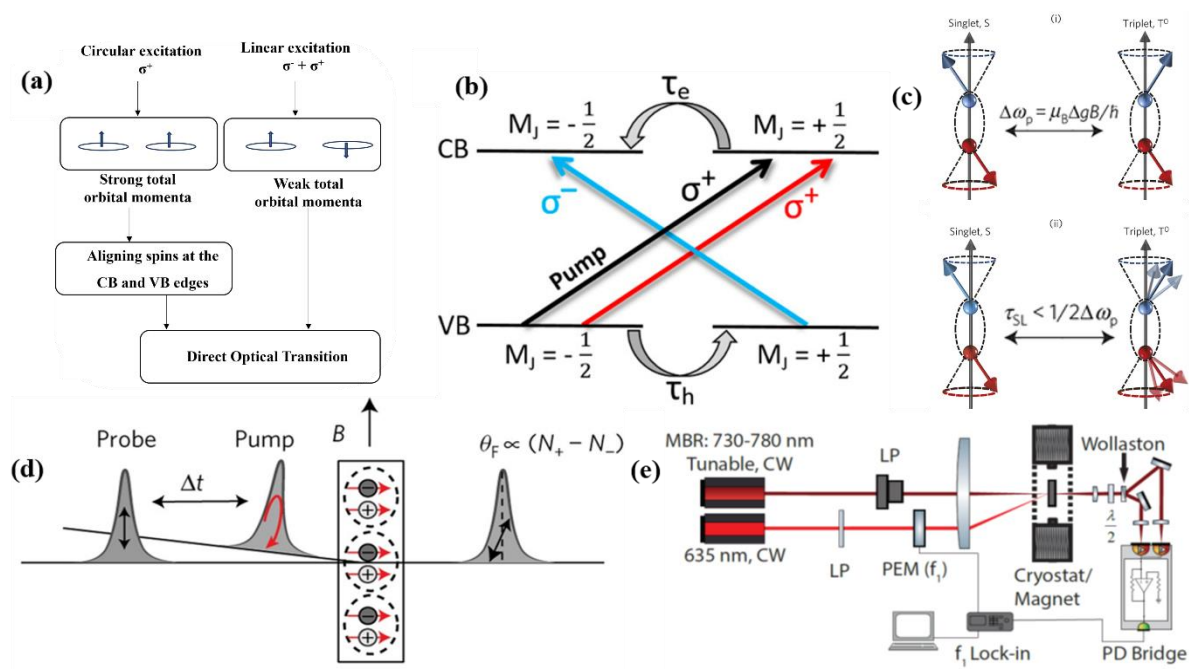

**FIG. 4**. (a) Schematic of the PL kinetics of halide perovskites under circular and linear polarized light excitation.[59] (b) Illustration of Faraday rotation measurement with the transversely applied magnetic field. Circularly polarized pump pulses produce spin-polarized excitons and quantized spin states. An exciton population imbalance ($N_\pm$) causes different light absorptions between RCP and LCP, and refraction indices ($\eta$), i.e., Faraday rotation $\theta_F \propto (\eta_+ - \eta_-) \propto (N_+ - N_-)$. Reproduced with permission.[127] Copyright © 2017 Springer Nature Limited. (c) Schematics of the Hanle experimental setup. Hanle measurements use the Voigt geometry and a split-coil electromagnet. 'Pump' optically orients spin-polarized excitons, while 'probe' analyses spin polarization via Faraday rotation. Half-wave plate and Wollaston beam splitter were used to detect the Faraday rotation. Reproduced with permission.[127] Copyright © 2017 Springer Nature Limited. (d) Illustration of optical transitions caused by circularly polarised light, where the pump excitation is denoted as the black arrow, and the probe pulses with helicities of σ⁺ and σ⁻ are represented in red and blue arrows, respectively. Reproduced with permission.[99] Copyright © 2020 American Chemical Society. (e) Schematic of two e-h spin-pair configurations, singlet (S) and triplet (T0), confined by a magnetic field (B) along the z-direction. Panels (i) and (ii) show the scenarios when the spin relaxation rate is lower and higher than the intersystem crossing (ISC) rate between S and $T^0$, respectively. A high spin relaxation rate causes incoherence of spin precession, diminishing the magnetic field effect. Reproduced with permission.[117] Copyright © 2015 Springer Nature Limited

As shown in **Fig. 4**(a), circular excitation generates left (σ+) or right (σ-) circularly polarized emission, while linear excitation combines excitons with opposite $J_z = +1$ and $J_z = -1$ states. In 3D bulk $MAPbI_3$, it has been reported that coherent SOC-induced momentum splitting of the CB and VB can align spins for circular excitation.[59] Although both CB and VB exhibit linear-k momentum splitting at the R point, the CB's splitting is more pronounced due to Pb's heavier mass than I atom and stronger SOC effect. This results in a slight indirect bandgap but predominantly behaves as a direct bandgap. However, in the case of linear excitation, it consistently leads to a direct optical transition. This coherent spin texture behaviour challenges previous understandings and suggests that both circular and linear excitations can induce direct optical transitions.

Furthermore, the SOC effect in $CsPbBr_3$ in the non-linear optical regime was investigated by Xu *et al.*, wherein the generation of transient spin-polarised electrons by two-photon absorption (TPA) using circularly polarized light excitation was demonstrated.[128] However, detecting high-order non-linear responses generated by TPA is unsuitable for analysis due to the poor detection limit. Therefore, THz emission spectroscopy with high sensitivity was employed to detect transient photocurrent in $CsPbBr_3$. The degree of spin polarisation of up to ~21.3 % was obtained at room temperature due to an asymmetric charge distribution in the Bloch states of CB and VB.

Spin-dephasing dynamics in halide perovskite materials is important due to its prominent function in quantum information processing applications. In a recent report, spin-dephasing dynamics in $CsPbBr_3$ were studied by time-resolved Faraday rotation (TRFR).[129] The optical spin polarisation and coherent spin precession at cryogenic and room temperatures were observed in $CsPbBr_3$ NCs. Two distinct spin-dephasing mechanisms were identified. At low temperatures, spin dephasing was mainly triggered by inhomogeneous hyperfine fields, which were suppressed by applying magnetic fields. At elevated temperatures, spin dephasing was caused primarily by thermally activated LO phonons.

The pump-probe techniques have become a prominent approach for time-resolved studies, widely applied in characterizing semiconductors, including halide perovskite materials. The exciton spin dynamics for perovskite materials can be investigated by different techniques such as pump-probe time-resolved Kerr rotation, pump-probe TRFR, and pump-probe time-resolved differential reflection.[99,127,130-133] **Fig. 4**(b) illustrates Faraday rotation measurement with the transversely applied magnetic field. A circularly polarised pump pulse populates excitons that are spin polarised along the beam path and in a coherent superposition of quantized spin states along the magnetic field (B). A population imbalance in the two exciton states ($N_\pm$) leads to different absorptions between left/right-handed circularly polarized (LCP/RCP, $\sigma^+/\sigma^-$) light and generates different indices of refraction ($\eta$), i.e., Faraday rotation $\theta_F \propto (\eta_+ - \eta_-) \propto (N_+ - N_-)$. A balanced photodiode bridge was used to measure the Faraday rotation, with a lock-in amplifier and an oscilloscope simultaneously monitoring the lock-in. The above circularly polarized 'pump' and linearly polarized 'probe' are combined in an ultrafast optical measurement of TRFR. As a function of time, the quantum beating between the quantized states causes an oscillatory Faraday rotation. The TRFR measurement utilizes a circularly polarized 'pump' (right or left-handed) to induce interband transition and generate spin-oriented $e^-$ and $h^+$ (spin-polarized exciton).[134] The linearly polarized 'probe' pulse is applied normal to the sample surface with a similar or different frequency to the pump. The magnetic field leads to the precession of the $e^-$/$h^+$ spin about $B$, with the Larmour frequency $\Omega = g_e \mu_B B/\hbar$. The induced spin coherence can be monitored from the Faraday rotation angle of the probe beam.

The optical Hanle effect is a steady-state 'pump-probe' measurement based on the Faraday rotation.[127] The Hanle effect was first observed in 1924.[135] Since then, it has been used to study spin relaxation, and lifetime.[136] **Fig. 4**(c) shows the setup of optical Hanle effect measurement, in which the samples were measured using CW lasers serving as 'pump' and 'probe'. The wavelength of the probe laser was tuned across the exciton band of the halide perovskite, while the pump laser had a fixed 635 nm wavelength. The measurements are set up in a Voigt geometry with a split coil electromagnet supplying the magnetic field. For measuring the Faraday rotation, the probe beam is adjusted to be at normal incidence on the sample, and the pump beam has a small angle of incidence through the same focusing lens as the probe beam. At the focal point of the pump beam, a polariser is positioned to allow or block any light based on its polarisation. A Wollaston polarising beam splitter and a half-wave plate are used to study and quantify the Faraday rotation. The optical Hanle effect measurement allows the probe beam wavelength to be precisely and independently tuned through the material's absorption edge with high energy resolution. The Faraday rotation signal is a time integration over all past time in this steady-state measurement, and it reduces with the application of a transverse magnetic field ($B$) because spin beating might repeal the time-integrated signal. The optical Hanle effect measurements use a circularly polarized 'pump' for the optical orientation of spin-polarised excitons and a linearly polarized 'probe' to measure exciton spin polarisation via Faraday rotation. Finally, a transverse applied magnetic field ($B$) diminishes the spin polarization. The Hanle curve obtained from this measurement gives general information about spin lifetime and recombination lifetime that can be extracted from the Lorentzian or Voigt function if the *g* factor is known.

The spin dynamics of charged carriers are usually probed using circularly polarized light. **Fig. 4**(d) shows the optical transition between the VB and CB using circularly polarised light.[99] When the pump pulse remains in positive helicity ($\sigma^+$), promoting carrier transition from VB to CB with angular momentum change from $M_{J, VB} = -1/2$ to $M_{J, VB} = +1/2$ (angular momentum conservation). The probe pulse is applied with a delay time, either in positive or negative helicity. TRPL and integrated PL have been utilized to study radiative recombination. Recently, Strohmair *et al.* explored the spin-related dynamics using DTS, which was utilized to examine the relaxation mechanisms, recombination dynamics, and spin lifetime of $CsPbI_3$ perovskite NCs at both room temperature and cryogenic temperatures.[99]

## 3.2 Magneto-optical techniques

Due to the magnetic nature of materials and the development of spintronic devices, the magnetic field effect (MFE) has been investigated to characterize magnetic materials.[137] Besides, MFE has also been explored to study various semiconducting materials, including halide perovskites.[117,138] Since applied magnetic fields can impact spin relaxation and net spin lifetime, this technique has been used to investigate various kinds of optoelectronic devices, such as solar cells and LEDs, which led to the developments of different analysis methods, including MEL, MPL, magneto-photoconductivity (MPC), and magneto-conductivity (MC).[139-143]

As illustrated in **Fig. 4**(e), the magnetic field (B) can modify the spin states by changing the inter-sublevel spin-mixing rates and in turn the spin sublevels in the spin-pair manifold. In MPL measurements, MFE can be observed when the spin relaxation rate is slower than the intersystem crossing (ISC), as shown in panel (i) of **Fig. 4**(e). On the other hand, the MFE diminishes when the spin relaxation is larger than the ISC rate [panel (ii) of **Fig. 4**(e)]. In halide perovskites, the spin-lattice relaxation rate is much slower than in typical semiconductors because of multiple factors, such as hyperfine interaction, SOC, and exchange interaction, leading to the observation of MFE. In the spin-pair species, the spintronic behaviour follows $\Delta\omega_p = \mu_B \Delta g B/\hbar$ ($\mu_B$ is the Bohr magneton); therefore, if the spin precession frequency difference between the spin-1/2 electron (e) and hole (h) is significant, the spin pair species promote ISC from the singlet, S, to triplet, $T^0$ state, which may occur several times throughout the spin-pair lifetime until the spin coherence is lost before achieving a steady state. The magnetic field B can alter the inter-sublevel spin mixing rates and recombination and dissociation rates for spin pairs on all sub-level spin configurations, leading to changes in PL/EL emission intensity and photocurrent. The presence of spin pair species leads to a spin mixing process in the spin sublevel, a necessary condition for MFE.[117]

## 4. Topological Insulating Phase in all-Inorganic Halide Perovskites

In 2012, through a high-throughput search, K. Yang et al. identified ternary halides, Cs(Sn,Pb,Ge)(Cl,Br,I)$_3$, as one of the previously unrealized families of topological insulator materials.[144] Since SOC is an essential component for the presence of nontrivial topological insulating phases, metal-halide perovskites with strong SOC have been suggested as potential materials for three-dimensional topological insulators in several independent studies.[145-150] The origin of nontrivial topological insulating character in the proposed all-inorganic metal-halide perovskites[145,147-150] is the same as that for prototypical three-dimensional topological insulator family $X_2Y_3$ (X = Bi, Sb; Y = Te, Se),[151-157] where topological bulk band gap is associated with the mass term in the Bernevig-Hughes-Zhang (BHZ) model.[158,159] In all-inorganic halide perovskite materials ABX$_3$, first-principles calculations[145] show that B 6p orbital states form the CB, while the VB maximum is formed by a linear combination of B 6s and X 5p orbital states with quenched orbital degrees of freedom in a singlet *s*-orbital symmetry. With even/odd parity of the basis sets constituting the VB/CB, band inversion and parity exchange lead to the nontrivial band topology. However, such a nontrivial topological insulating phase is not intrinsic like $X_2Y_3$ family and is generally induced rather by external stimuli.

As shown in **Fig. 5**(a), for all-inorganic metal halide perovskites $ABX_3$, the effective 4 × 4 continuum Hamiltonian, up to quadratic power in momentum $k$ in the vicinity of R-point, can be written in the subspace spanned by spin doublets (s = ½) forming VB and total angular momentum doublets (j = ½) forming CB as[145]

$$H_{eff}(\vec{k}) = \varepsilon_0(\vec{k})\tau_0 \otimes \sigma_0 + m(\vec{k})\tau_z \otimes \sigma_0 + \Delta \tau_y \otimes (\vec{k} \cdot \vec{\sigma}^*) \tag{4.1}$$

where Pauli matrices $\tau_{i=x,y,z}$ and $\sigma_{i=x,y,z}$ denote orbital and spin degrees of freedom, respectively, $m(\vec{k}) = [\xi_s(\vec{k}) - \xi_p(\vec{k}) + \lambda]/2$ are effective on-site potentials, and $\Delta = 2t_{sp}^\sigma/\sqrt{3}$ is the magnitude of effective inter-orbital hopping. Here

$$\xi_s(\vec{k}) = \varepsilon_s + 6t_{ss}^\sigma - \left[ t_{ss}^\sigma + \frac{8}{3} \frac{(t_{sp}^\sigma)^2}{\varepsilon_p - \varepsilon_s - 2t_{pp}^\sigma - 4t_{pp}^\pi - 6t_{ss}^\sigma + \lambda/2} \right] k^2 \tag{4.1a}$$

$$\xi_p(\vec{k}) = \varepsilon_p - 2t_{pp}^\sigma - 4t_{pp}^\pi + \frac{1}{3}\left[ t_{pp}^\sigma + 2t_{pp}^\pi \right] k^2 \tag{4.1b}$$

where $\varepsilon_s$ and $\varepsilon_p > \varepsilon_s$ are on-site potentials, $\lambda$ is the strength of on-site SOC, while $t_{pp}^\sigma, t_{ss}^\sigma, t_{pp}^\pi$ and $t_{sp}^\sigma$ are intra- and inter-orbital Slater-Koster hopping integrals, respectively. The effective Hamiltonian $H_{eff}(\vec{k})$ $\varepsilon_0(\vec{k})$ breaks the particle-hole symmetry while $m(\vec{k})$ induces mass in the bulk band spectrum. As a result, similar to the three-dimensional topological insulator family $X_2Y_3$, where the Rashba SOC simulates the surface/interface states,[151,154] this isotropic massive Dirac Hamiltonian characterizes the band topology of metal-halide perovskites through sign change of the mass term at $\vec{k} = 0$. That is, considering that $\varepsilon_p > \varepsilon_s$ and all the hopping parameters are positive, the system remains a trivial insulator when $\varepsilon_s + 6t_{ss}^\sigma + \lambda < \varepsilon_p - 2t_{pp}^\sigma - 4t_{pp}^\pi$, while becomes a topological insulator when $\varepsilon_s + 6t_{ss}^\sigma + \lambda > \varepsilon_p - 2t_{pp}^\sigma - 4t_{pp}^\pi$. The positive mass term or negative band gap condition for the topological insulating phase can be written in terms of half of the VB and CB width, $W_s/2 = 6t_{ss}^\sigma$ and $W_p/2 = 2t_{pp}^\sigma + 4t_{pp}^\pi$, respectively, as

$$\frac{W_s + W_p}{2} + \lambda > \Delta\varepsilon, \tag{4.2}$$

where $\Delta\varepsilon \equiv \varepsilon_p - \varepsilon_s$. It implies that ABX$_3$ enters the topological insulating phase when the sum of SOC splitting, and the half bandwidth of the VB and CB exceeds the on-site potential difference between *s* and *p* orbital states on the B-site cation and X-site anions. Obtained from first principle calculations and effective continuum model[145], energy configuration of effective states near the Fermi level, SOC and pressure-induced band inversion between the $|s = 1/2\rangle$ and $|j = 1/2\rangle$ states, sign change of mass term or band gap associated with the lattice constant, and topological phase transition between the trivial and nontrivial phase of metal-halide perovskites ABX$_3$ are displayed in **Fig. 5**(b) and **5**(c).

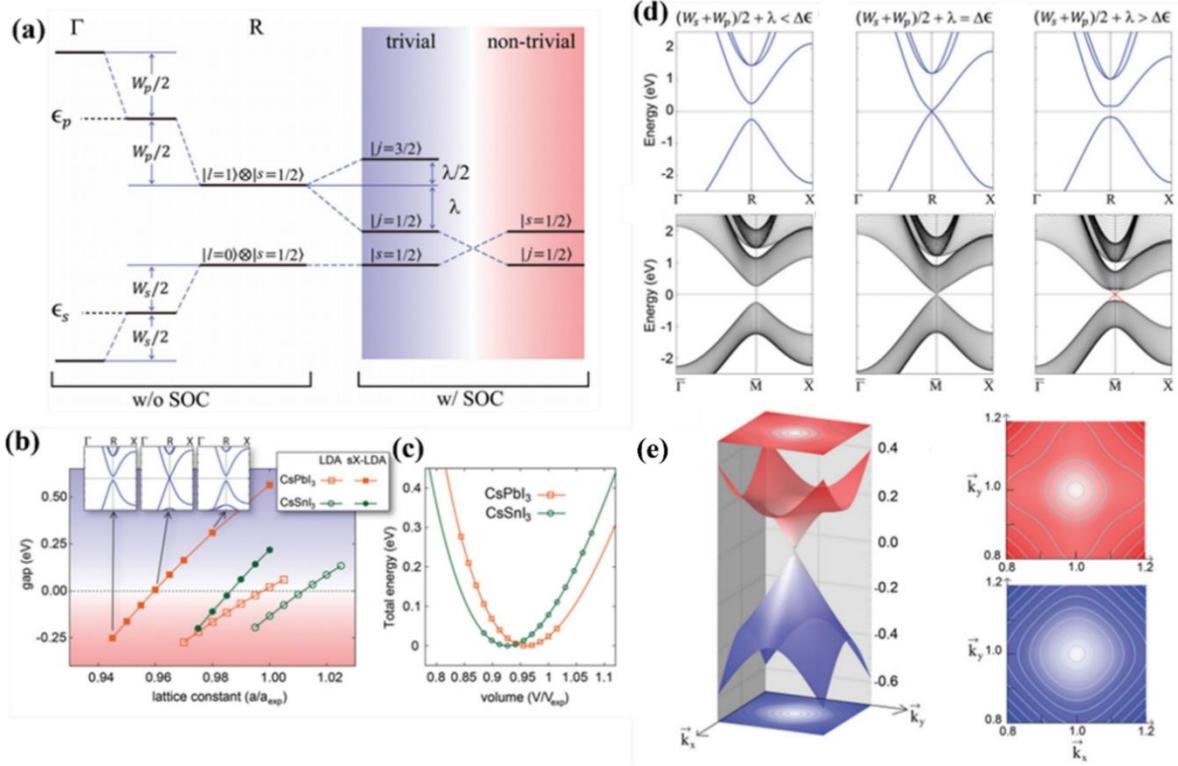

**FIG. 5.** (a) Basic electronic structure, band inversion, and topological phase transition in all-inorganic metal-halide perovskites $ABX_3$. SOC and hydrostatic pressure operation induce band inversion, and the system enters a topologically nontrivial insulating phase. (b) Dependence of the band gap value sign on the lattice constants. (c) Total energy dependence of the unit-cell volume for $CsPbI_3$ and $CsSnI_3$. The positive band gap values (negative mass term; $m(\vec{k}) < 0$) represent the topologically trivial phase, whereas the negative band gap values (positive mass term; $m(\vec{k}) > 0$) represent the topologically nontrivial phase. (d) Tight binding calculations showing the topological phase transition in the bulk band spectrum (upper row) and (0,0,1) surface spectrum (lower row) for 51-layer slab geometry with topologically protected surface states. (e) Three-dimensional band dispersion for gapless surface states and its isoenergy contour map. Reproduced with permission.[145] Copyright ©2012 American Physical Society.

The constraint set by microscopic parameters on the nontrivial character of band topology suggests that a topologically nontrivial insulating phase can be achieved in $ABX_3$ halide perovskite materials with large SOC and large hopping parameters. The nontrivial insulating phase can be realized either with heavy metal elements on B-sites and heavy halogen elements on X-site or via an externally controlled environment, such as pressure and strain-induced enhancement in the bandwidth by decreasing the lattice constants. For instance, through various independent studies on $CsBI_3$ (B = Pb, Sn)[145], $CsBX_3$ (B = Pb, Sn; X = Cl, Br, I)[148], and $CsSnX_3$ (X = I, Br, Cl)[149], it has been predicted that all-inorganic $ABX_3$ halide perovskite compounds display topologically nontrivial insulating phase under reasonable hydrostatic pressure. For instance, pressure-induced variation in the band gap of $CsBX_3$ (B = Pb, Sn; X = Cl, Br, I), red-shift in the trivial band gap while blue-shift in the nontrivial band gap with increase in pressure, demonstrates a topological phase transition from trivial to a nontrivial phase.[148] Furthermore, topological insulating phase in $CsPbI_3$ with ferroelectric response [147] and $CsPbBr_xI_{3-x}$ ($x$ = 0, 1, 2, and 3) mixed halide perovskite compound[160] has been discussed under accessible strain. Consistent with ferroelectric all-inorganic metal halide perovskite $CsPbI_3$,[147] a similar structure of low energy valence and conduction Rashba bands, formed by spin and total angular momentum doublets $S = 1/2$ and $J = 1/2$, has been found in organic-inorganic hybrid metal halide perovskites, $AMX_3$ where A=$CH_3NH_3$, i.e., methylammonium (MA); M = Pb and Sn; and X = I and Br.[161] It indicates that an electric field switchable $S = 1/2$ and $J = 1/2$ Rashba bands, as predicted for organic-inorganic hybrid metal halide perovskites[161], may also be a key ingredient for spintronic applications in all-inorganic metal halide perovskites with nontrivial band topology and ferroelectric response.

Unique electrical and optical properties in the topological insulating phase, due to the surface states intertwining with the bulk band topology, give rise to substantial implications in the field of electronics, spintronics, plasmonics, and optoelectronics. For instance, topological insulators materials have the ability to overcome Boltzmann tyranny and lead to low-power topological quantum electronic device applications owning to the Rashba effect[162], negative capacitance effect[163], and quantum confinement effects[164]. In addition, owning to the spin-momentum locked helical surface/edge states, it has been theorized that topological insulator materials have the potential to enable novel topological spintronic device applications.[165-173] Furthermore, owing to the fact that surface/edge states in the topological insulators exhibit an exotic electronic response to light, significant progress has also been made in exploring the optoelectronic/plasmonic properties and device applications with topological insulator materials. Notable examples include but are not limited to, topological quantum phase transitions [174], surface plasmons for energy-harvesting applications[175,176], wide bandwidth photodetection from terahertz to the infrared[177], transparent conductive electrodes with topological insulator nanostructures[178], ultrafast photo-response with topological insulator $Bi_2Te_3$ nanowires[179], high light responsivity, high detectivity, and a fast response speed in the topological insulator $Bi_2Se_3$/Si heterostructure[180], and control over spin-polarized photocurrents with helicity or linear polarization of light,[181,182] leading to novel spin-optoelectronic or opto-spintronic devices[183].

Since the origin and the tunability of band topology in all-inorganic halide perovskites is like that in the topological insulator family $X_2Y_3$ (X = Bi, Sb; Y = Te, Se), such intriguing optoelectronic properties displayed by the $X_2Y_3$ family provide a hint of band topology driven high-performance optoelectronic device applications in all-inorganic halide perovskites. Along with excellent power conversion efficiency, tunable physical properties, cost-effective synthesis, and long-term thermal and environmental stability, strong SOC and band topology may push the boundary even further for the novel applications of all-inorganic halide perovskites.

## 5. Summary and Outlook

In summary, the SOC effect in all-inorganic halide perovskite has been systematically discussed, along with its influence on the band structure, charge recombination, light emission, spin lifetime and other properties. While possessing better stability compared to the hybrid counterparts, all-inorganic halide perovskites retain heavy Pb atoms, which leads to a strong SOC effect that exerts a significant influence over their physical properties and device performance. In addition, the inorganic cation Cs possesses a relatively high symmetry compared to the organic ones, which underlies the composition-property correlation in this class of halide perovskite materials. However, both internal and external factors may still induce structural inversion asymmetry and slightly distort the metal-halide octahedra, thereby inducing the Rashba effect and the band splitting.

Recent investigations in the domain of all-inorganic halide perovskites have unveiled several intriguing developments. There have been seminal works revealing the relative positions between the bright triplet and dark singlet states.[41-43,107,108] Furthermore, a debate persists regarding the existence of the Rashba effect in these materials, prompting a quest for definitive experimental evidence.[50] This ongoing debate highlights the need for further research to elucidate the intricate interplay between inversion asymmetry, lattice dynamics, defect characteristics, and SOC, offering prospects for advancing Rashba engineering and the control of quantum materials.

The research exploring the correlation between spin lifetime and SOC is still in its infancy. Different techniques such as transient absorption spectroscopy, TRFR, TPA, DTS, and TRPL were used to measure all-inorganic halide perovskites with different compositions under different conditions, thus it is challenging to consolidate the results. All-inorganic halide perovskite is hypothesized to potentially exhibit a static Rashba effect at temperatures below 60 K and a dynamic Rashba effect at higher temperatures.[93] It is of pivotal importance to explore the dynamic Rashba contribution arising from the thermally activated polar distortion of $PbX_6$ octahedra, as this might offer a means to control spin lifetime and other properties at application-relevant temperatures.

Topological insulating phases, with unique electrical and optical properties, hold the promise of advancing electronics, spintronics, and optoelectronics, enabling low-power quantum devices. The adaptable band topology in all-inorganic halide perovskites offers high-performance optoelectronic potential, driven by efficiency, tunability, stability, SOC, and suggests that the band topology could be the next frontier for the research on all-inorganic halide perovskites. It has been predicted that the halide perovskite compounds exhibit a red shift in trivial bandgap while a blue shift in nontrivial bandgap by increasing pressure.[148] The halide perovskite-based 3D topological insulators can offer a new avenue for diverse potential applications by introducing a new degree of freedom, the so-called topological order. The topological surface states protected by time-reversal symmetry in 3D cubic halide perovskites might offer new physics, and the topological perovskite/perovskite heterostructure interface could serve as a playground where diverse types of quasiparticles interact with each other in the confined 2D region and may exhibit exotic properties. Consequently, further experimental, and theoretical research should be carried out to explore the topological order in other halide perovskite systems with different A- and B-site cations.

The exciting properties enabled by the hidden force of SOC will propel all-inorganic halide perovskite materials to become a rising star in exploring novel applications. Metal halide perovskites hold great promise in revolutionizing emerging disciplines such as spintronics and opto-spintronics due to their unique features, like low-cost fabrication, tunable band gaps, and exceptional charge carrier properties. From their robust SOC to their applications in spintronic devices like vertical spin-valve configurations, spin-LEDs, and spin photovoltaic devices, these materials have already demonstrated their potential, and have open pathways for future novel devices. The ongoing exploration of inorganic metal halide perovskites and refinement of their properties offer opportunities for improved device performance and the potential to advance next-generation optoelectronics, spintronics, opto-spintronics, and related fields.

**Acknowledgements**

This work was supported by the UNSW SHARP Project (RG163043), Australian Research Council (DP190103316 and DP230101847) and the Hong Kong Global STEM Professorship. Simrjit Singh acknowledges funding from the European Union's Horizon 2020 research and innovation programme under the Marie Skłodowska-Curie grant agreement No 899987. Pankaj Sharma acknowledges support from a Flinders University start-up grant.

**Competing Interests**

The authors declare no competing interests.

**Data Availability**

Data sharing is not applicable to this article as no new data were created or analyzed in this study.